\documentclass[ios]{emulateapj}

\usepackage{graphicx}
\usepackage{color}
\usepackage{amssymb}
\usepackage{amsmath}
\usepackage{hyperref}
\newcommand{\nustar}{{\it NuSTAR}}

\newcommand{\gs}{GS~0834$-$430}

\def\arcsec{$\,^{\prime\prime}$}

\def\simless{\mathbin{\lower 3pt\hbox
   {$\rlap{\raise 5pt\hbox{$\char'074$}}\mathchar"7218$}}} 
\def\simgreat{\mathbin{\lower 3pt\hbox
   {$\rlap{\raise 5pt\hbox{$\char'076$}}\mathchar"7218$}}} 

\slugcomment{Accepted to The Astrophysical Journal}

\shorttitle{Hard Lags in GS~0834$-$430}
\shortauthors{Miyasaka et al.}

\begin{document}


\title{\nustar\ detection of Hard X-ray Phase Lags from the Accreting Pulsar \gs}

\author{Hiromasa Miyasaka\altaffilmark{1}}
\author{Matteo Bachetti\altaffilmark{2,3}}
\author{Fiona A. Harrison\altaffilmark{1}}
\author{Felix F\"urst\altaffilmark{1}}
\author{Didier Barret\altaffilmark{2,3}}
\author{Eric C. Bellm\altaffilmark{1}}
\author{Steven E. Boggs\altaffilmark{4}}
\author{Deepto Chakrabarty\altaffilmark{5}}
\author{Jerome Chenevez\altaffilmark{6}}
\author{Finn E. Christensen\altaffilmark{6}}
\author{William W. Craig\altaffilmark{4,7}}
\author{Brian W. Grefenstette\altaffilmark{1}}
\author{Charles J. Hailey\altaffilmark{8}}
\author{Kristin K. Madsen\altaffilmark{1}}
\author{Lorenzo Natalucci\altaffilmark{9}}
\author{Katja Pottschmidt\altaffilmark{10}}
\author{Daniel Stern\altaffilmark{11}}
\author{John A. Tomsick\altaffilmark{4}}
\author{Dominic J. Walton\altaffilmark{1}}
\author{J\"orn Wilms\altaffilmark{12}}
\author{William Zhang\altaffilmark{13}}

\altaffiltext{1}{Cahill Center for Astronomy and Astrophysics, California Institute of Technology, Pasadena, CA 91125; miyasaka@srl.caltech.edu.  }
\altaffiltext{2}{Universit\'e de Toulouse; UPS-OMP; IRAP; Toulouse, France}
\altaffiltext{3}{CNRS; Institut de Recherche en Astrophysique et Plan\'etologie; 9 Av. colonel Roche, BP 44346, F-31028 Toulouse cedex 4, France}
\altaffiltext{4}{Space Sciences Laboratory, University of California, Berkeley, CA 94720, USA}
\altaffiltext{5}{Kavli Institute for Astrophysics and Space Research, Massachusetts Institute of Technology, Cambridge, MA 02139}
\altaffiltext{6}{DTU Space, National Space Institute, Technical University of Denmark, Elektrovej 327, DK-2800 Lyngby, Denmark}
\altaffiltext{7}{Lawrence Livermore National Laboratory, Livermore, CA 94550, USA}
\altaffiltext{8}{Columbia Astrophysics Laboratory, Columbia University, New York, NY 10027, USA}
\altaffiltext{9}{Istituto di Astrofisica e Planetologia Spaziali, INAF, Via Fosso del Cavaliere 100, Roma I-00133, Italy}
\altaffiltext{10}{CRESST, UMBC, and NASA GSFC, Code 661, Greenbelt, MD 20771,
USA}
\altaffiltext{11}{Jet Propulsion Laboratory, California Institute of Technology, Pasadena, CA 91109, USA}
\altaffiltext{12}{Dr. Karl-Remeis-Sternwarte and ECAP, Sternwartstr. 7, 96
049 Bamberg, Germany}
\altaffiltext{13}{NASA Goddard Space Flight Center, Greenbelt, MD 20771, USA}

\begin{abstract}
The \nustar\ hard X-ray telescope observed the transient Be/X-ray binary \gs\ during its 2012 outburst -- the first active state of this system observed
in the past 19 years.   We performed timing and spectral analysis, and measured the X-ray spectrum between 3 -- 79\,keV with high statistical significance.
We find the phase-averaged spectrum to be consistent with that observed in many other magnetized accreting pulsars.  We fail to 
detect cyclotron resonance scattering features that would allow us to constrain the pulsar's magnetic field in either 
phase-averaged or phase-resolved
spectra. 
Timing analysis shows a clearly detected pulse period of $\sim12.29$\,s in all energy bands. The pulse profiles show a strong, energy-dependent hard phase lag  of up to 0.3 cycles in phase, or about 4s.
Such dramatic energy-dependent lags in the pulse profile have never before been reported in high-mass X-ray binary (HMXB) pulsars.  Previously reported
lags have been significantly smaller in phase and restricted to low-energies (E$<$10\,keV). 
We investigate the possible mechanisms that might produce this energy-dependent pulse phase shift. 
We find the most likely explanation for this effect to be a complex beam geometry.
\end{abstract}


\section{Introduction}

The transient X-ray source \gs\ was discovered  in outburst by the {\em Granat}/WATCH  X-ray sky monitor
in 1990 \citep[the source was named GRS~0831$-$429]{Sunyaev90}.    Subsequent observations by  {\em Ginga}  \citep{ade+1992},
{\em ROSAT} \citep{Belloni+93} and {\em Granat}/ART$-$P \citep{1991IAUC.5180....1S} identified a 12.3\,s period pulsar with a low
($\sim$10\%) pulsed fraction.   The
BATSE all-sky monitor on the {\em Compton Gamma-Ray Observatory} detected 7 outbursts from \gs\
over the period 1991 April -- 1993 June  \citep{Wilson+97}.   The first five occurred at a regular spacing of about 106 days, while the 
last two were spaced by about 140 days.    Using the first five BATSE-detected outbursts,
\citet{Wilson+97} determined
1-$\sigma$ limits on the orbital period of $P_{\rm orb} = 105.8 \pm 0.4$\,d and on the eccentricity of $0.10 < e < 0.17$.
Subsequently, the source became dormant for a period of 19 years. 

The hard spectrum, the recurring nature of the outbursts, and the presence of pulsations suggested that \gs\ is
a Be/X-ray binary with a magnetized neutron star accreting from the dense equatorial wind of its companion \citep{Belloni+93}. 
A low orbital inclination could explain the low pulsed fraction, assuming that the spin and orbital angular momentum vectors
are aligned \citep{Wilson+97}.   The low orbital eccentricity and the combination of regular and irregular outbursts of similar intensity are unusual for such systems. \citet{Israel+00} identified the optical counterpart, most likely a B0--2 V--IIIe spectral type star located at a distance of $\sim5$\,kpc, confirming the Be-star/X-ray binary nature of \gs.

The initial observations noted unusual features in the pulse profile as a function of energy.
Both \citet{ade+1992} and \citet{Wilson+97} reported a variation of the pulse shape with energy, 
with a primary and a secondary peak dominant in different energy bands. 
\citet{Wilson+97} remarked on the presence of some shift between profiles in different energy bands,
but did not discuss it further.   Significant phase shifts as a function of energy in accreting pulsars are rare, and this singles
out \gs\ as an interesting object for phase-resolved spectroscopy.      

On UT June 26, 2012 {\em INTEGRAL} \citep{integral03}, during one of its Galactic Plane Scans, discovered that \gs\ was in outburst for the first time in 19 years,
with a 18 -- 60\,keV flux of 109\,mCrab \citep{2012ATel.4218....1D}. {\em Swift} \citep{swift05} measured an absorption column of $2.2^{+2.63}_{-1.66}\times10^{22}$\,cm$^{-2}$ \citep{2012GCN..13388...1B}, and {\em Fermi}/GBM \citep{fermigbm09} detected pulsations with barycentric frequency 81.3357(4)\,mHz on UT June 28--30, and 81.3377(3)\,mHz on UT June~30--July 2
\footnote{See full period evolution, as detected by GBM, at
http://gammaray.nsstc.nasa.gov/gbm/science/pulsars{\slash}lightcurves{\slash}gs0834.html
} \citep{2012ATel.4235....1J}.

\begin{figure}[htbp] 
   \centering
   \includegraphics[width=\columnwidth]{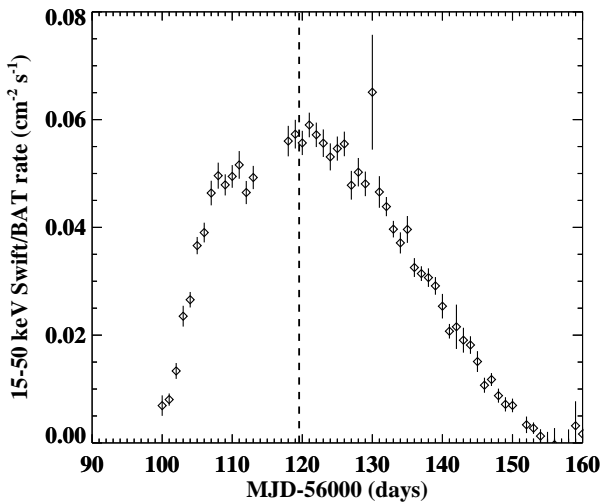} 
   \caption{The {\em Swift}/BAT light curve in the 15 -- 50~keV range for GS 0834$-$430 during the
2012 July outburst.  The vertical dashed line marks the midpoint of the
\nustar\ observation, which occurred at the very peak of the outburst.
}
   \label{fig:bat}
\end{figure}

On UT July 11, at the peak of the outburst (see Figure~\ref{fig:bat}), the \textsl{Nuclear Spectroscopic Telescope Array} \citep[\nustar,][]{nustar13}
 observed \gs\ for about 30\,ks.   
In this paper we present these data, including the first phase-resolved spectroscopy simultaneously 
spanning the entire energy range from 3 -- 79\,keV.

\section{Observations and Data Reduction}

\nustar\ is the first focusing high energy X-ray telescope in space, sensitive between 3--79\,keV. Two grazing incidence optics focus X-rays on two independent CZT detectors, called Focal Plane Modules A and B (FPMA and FPMB). For a detailed description of the instrument see \citet{nustar13}.

\nustar\ observed \gs\ on 2012 July 11 from 02:40:00 -- 23:40:00 UT as part of its observatory commissioning program.    
The total exposure time, corrected for dead time, Earth occultation
and South Atlantic Anomaly (SAA) passages, was 31.0\,ks.  The total dead time-corrected count rate was $52.01 \pm 0.05$\,cts\,s$^{-1}$ (3 -- 79\,keV) in each of the
two co-aligned telescope modules.  For this observation the source was placed near the
optical axis, on detector 0 \citep[see][for details of the focal plane layout]{nustar13}. 
  Individual photons are time-tagged to a relative accuracy of 2\,$\mu$s, and absolute (relative to UT)
calibration of the on-board clock is good to 2\,ms rms.    The spectral resolution ranges from 400\,eV (FHWM) at 10
\,keV to 1\,keV (FWHM) at 79\,keV.

We reduced and analyzed the data using the {\em NuSTAR Data Analysis Software (NuSTARDAS)} v1.1.1 and CALDB
version 20130509.   The data were filtered for intervals of high background due to SAA passages.    
To extract data for spectral analysis we used a circular aperture of radius 65\arcsec\ centered on the source position.
We determined the background by extracting counts from a region 100\arcsec\ in radius on a different detector than the
source, since \gs\ is bright and the PSF wings contaminate most of detector 0.
At 10\,keV the source is 10$^3$ times brighter than the background, and at 80\,keV the source is still 10 times above the
background, so small systematics uncertainties due to choosing a different chip for background subtraction are unimportant. We rebinned the spectrum in order to have at least 50 counts per channel.

\section{Pulse Profile Analysis}

\subsection{Pulse period determination}

\begin{deluxetable}{lcc}
\tablecolumns{3}
\tablewidth{0pc}
\tabletypesize{\scriptsize}
\tablecaption{Ephemeris for GS0834-430 \citep{Wilson+97}\label{tab:ephemeris}}
\startdata
\hline
Projected semi-major axis                    &     $a_x \sin i$  [$lt-s$]                     &   $128^{+47}_{-38}$      \\
Orbital period at the epoch                   &     $P_{\rm orb}$ [days]                      &   $105.8\pm0.4$          \\
Eccentricity $ \times$    $a_x \sin i$         &     $e\ a_x\sin i$  [$lt-s$]                 &  $15.3^{+3.3}_{-0.9}$      \\  
Longitude of periastron                        &     $\omega$ [degree]     &  $140^{+35}_{-53}$      \\     
Epoch for mean longitude of 90 deg   &    $T_{90}$  [MJD]                            &  $48809.5$          
\enddata
\end{deluxetable}

In order to determine the pulse period of \gs, we first corrected the photon arrival times to the solar system barycenter using the HEASARC FTOOL {\tt  barycorr}, with the default DE$-$200 solar system ephemeris, adopting the sky position measured by {\em ROSAT} (\citealt{Belloni+93}; $\alpha=8^h$35$^m$55$^s$.1, 
$\delta=-43^{\circ}$11$'$22$''$, J2000) and the orbital ephemeris determined
by \citet{Wilson+97}, which is given in Table~\ref{tab:ephemeris}.  We determined the pulse period of \gs\   using an epoch folding technique \citep{Leahy+83}. 
A Z$^2_3$ analysis \citep{Buccheri+83} yields a period of $12.293(1)$\,s.
This period is consistent with the range measured by GBM.    We note the period is not stable due to the combination of spin-down
and spin-up due to accretion torque \citep{DavidsonOstriker73}, but does not change significantly over the \nustar\ observation. We measure a pulsed fraction, defined as the ratio of the half amplitude of the pulse to the mean count rate, of 15.5(4)\% over the whole energy range, going up to 18.3(8)\% in the 40 -- 80\,keV range.
This is comparable to the BATSE measurement  \citep{Wilson+97} of $\lesssim$15\% in the 20 -- 50\,keV range.  
Figure 12 from that work also shows best fit values of $\sim$15--40\% in the 50 -- 70\,keV band for several BATSE 
observations, providing marginal evidence for an increase in the pulse fraction with energy.

\subsection{Energy-dependent pulse profile}

\begin{figure}[htbp]
   \centering
   \includegraphics[width=\columnwidth]{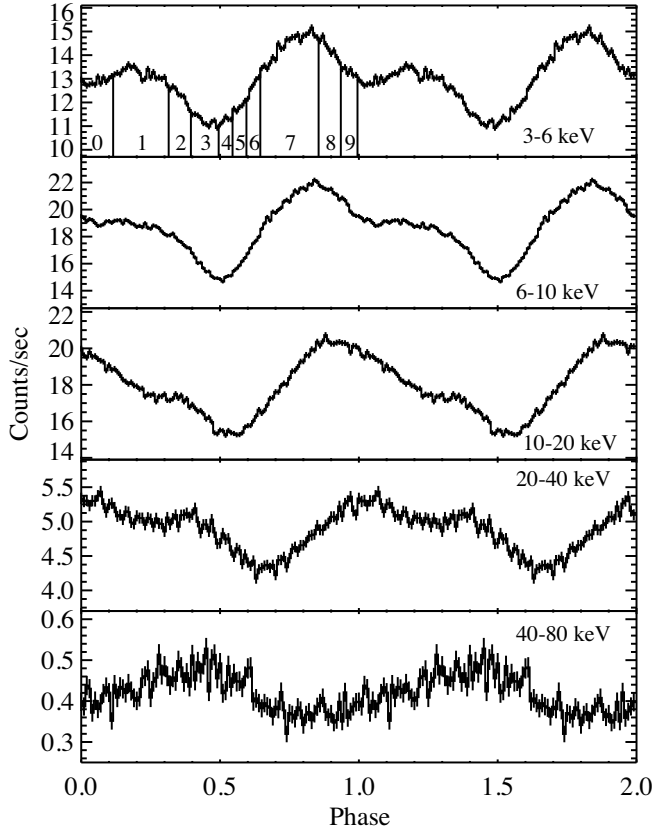} 
   \caption{The pulse profile in four different energy bands.   A strong dependence with energy is clearly visible. Intervals in the top pulse profile show the phases used for phase-resolved spectroscopy (see \S\ref{sec:phaseres}).}
   \label{fig:countrate}
\end{figure}

Figure~\ref{fig:countrate} shows the pulse profile in five different energy bands over phase range 0 to 2 (two pulse cycles). Phase 0 is chosen to be at
MJD 56119.203555 (TDB).     The pulse profile is clearly stable, and strongly dependent on energy.  Two peaks are visible; a secondary (smaller) peak in the 3 -- 6\,keV band at phase $\sim$0.2, and a larger peak at phase $\sim$0.8.         
It is also clear that the phase of the larger peak shifts significantly with increasing
X-ray energy, such that in the 20 -- 40\,keV band it has shifted in phase by 0.3 cycles (peaking at phase 0.1). Such a large and dramatic energy dependence has not been previously observed in any accreting X-ray pulsar. In the 40 -- 80\,keV range  the secondary peak becomes dominant (peaking at phase $\sim$1.4).   

\begin{figure}[htbp]
   \centering
   \vspace{0.5cm}
  \includegraphics[width=\columnwidth]{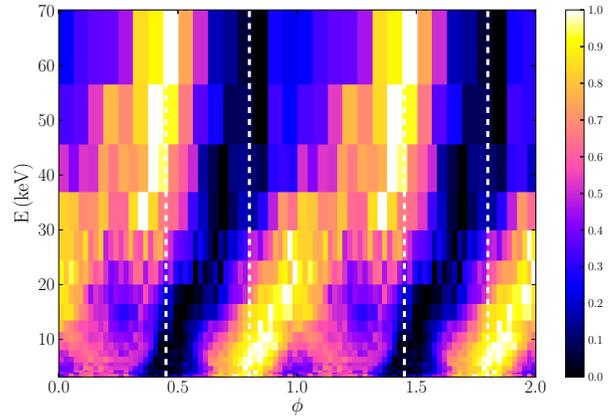}
   \caption{Evolution of the DC level-subtracted pulse profile with energy. We started from a map where every profile was calculated in an energy band of $\sim$0.4\,keV and was divided in 48 bins. In order to attain a better S/N at high energy, we then averaged over an increasing number of bins at higher energies, following approximately a geometrical progression both in phase and in energy. Each profile is background-subtracted and normalized from 0 to 1. This normalized rate is color-coded, ranging from the lowest count-rates in black to the highest in white. The profiles are repeated twice in phase for clarity. The white dashed lines indicate roughly the phase of the peak and the dip, respectively, at the softest energies to guide the eye. The image is not smoothed, and the noise at high energy is due to the poor statistics.
   The minimum and the peak of the profiles are shifting in phase with energy, showing a clear hard lag.}
   \label{fig:pulsemap}
\end{figure}

\begin{figure}[htbp]
   \centering
   \vspace{0.5cm}
  \includegraphics[width=\columnwidth]{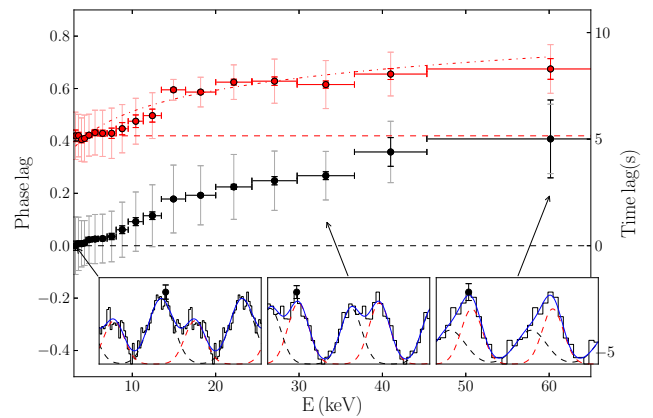}
   \caption{Phase lags of the two peaks, as obtained by fitting each rebinned profile in Fig.~\ref{fig:pulsemap} with two independent Gaussians, relative to the phase of the first peak in the lowest energy band. Black is the main peak, red the secondary. Full-color error bars are 1-$\sigma$ uncertainties. Light-colored error bars indicate the HWHM of the Gaussians. Dashed lines show the phase of the two peaks in the lowest energy band.  The dash-dotted line indicates the best fit of equation~\ref{eq:compt} to the secondary peak lags (see also \S\ref{sec:discussion}). We chose the secondary peak because it is better constrained over the whole energy range. The insets show the fit in the first energy band, at $\sim$35\,keV where the secondary peak becomes dominant, and the last energy band, respectively.}
   \label{fig:lagfit}
   \vspace{0.1in}
\end{figure}

Figure~\ref{fig:pulsemap} shows a map of the DC level-subtracted pulse profile as a function of energy. The binning is variable over the map, in order to attain a better S/N as explained in the figure caption.
Focusing on the strongest feature which peaks at phase 0.8 in the softest bands, there is a clear and almost linear shift of the pulse peak with energy until phase $\sim$1.1. 
To better investigate the behavior of the profile with energy, we fitted every average profile with a double Gaussian. The result of this procedure is shown in Figure~\ref{fig:lagfit}. The two peaks move in phase in a quasi-parallel fashion below $\sim$30\,keV, where the first peak is dominant over the second. 
Around 35\,keV, the secondary peak becomes dominant.  At higher energies, 
the primary peak becomes wider and less significant, and the behavior is less
clear.  The only evidence for the presence of the primary peak is the fact that
the high-energy profile has different slopes on the two sides, providing a hint
of a small contribution from the fading primary peak.

\section{Phase-averaged Spectrum}\label{sec:spec}

\begin{figure}[htbp]
   \centering
  \includegraphics[width=\columnwidth]{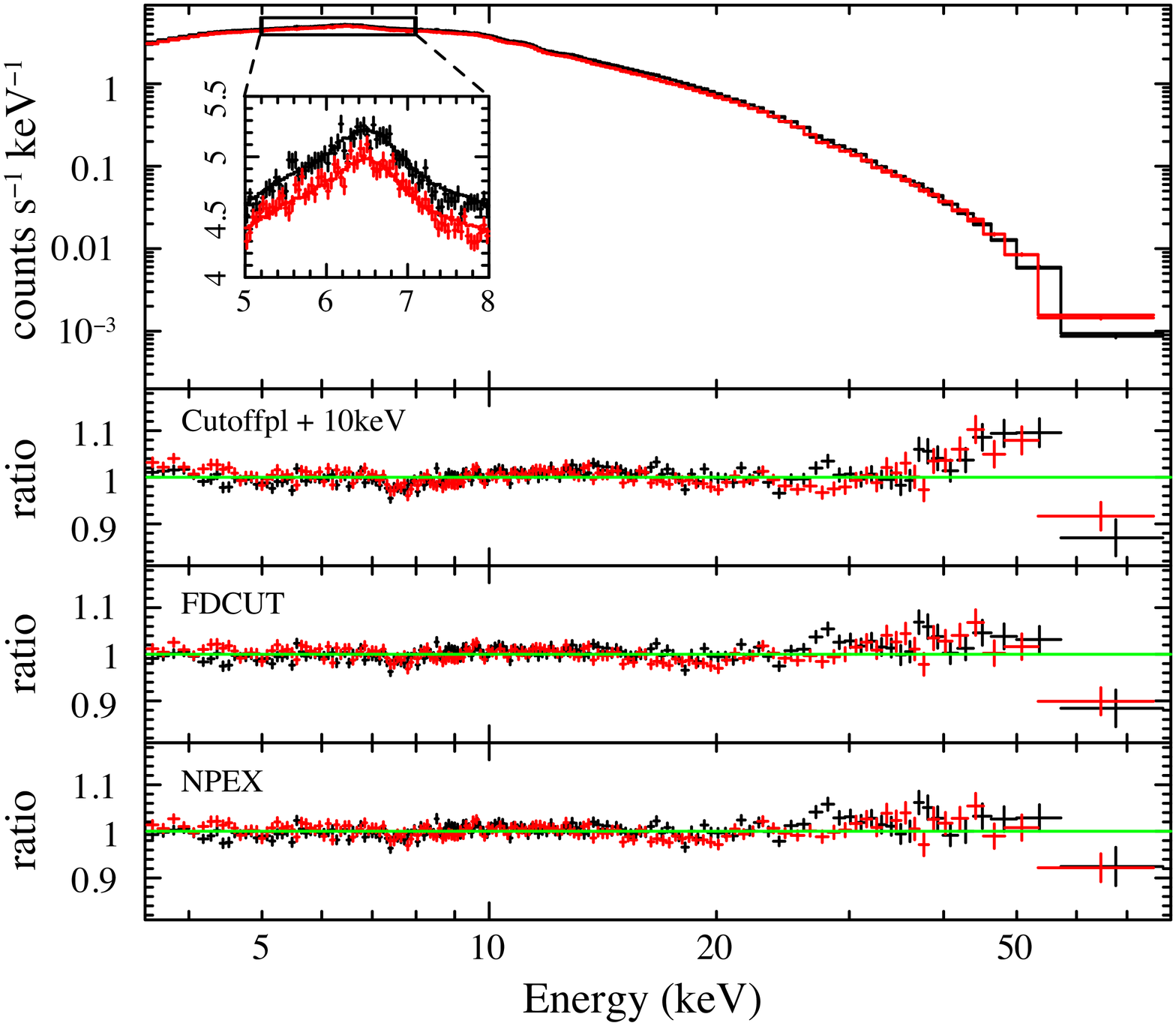}
   \caption{Phase-averaged spectrum fit using the {\tt NPEX}  continuum model. Black points represent data from module FPMA, 
   and the red are from FPMB. 
   The top panel shows the counts spectrum, with the data-to-model ratio for CutoffPL+10keV, FDCUT and NPEX 
   shown on the second to fourth panels, respectively. The region around the iron line is shown in the inset.}
   \label{fig:fdcut}
\end{figure}

The phase-averaged X-ray spectrum of \gs\ is broadly similar to that seen in many X-ray binary pulsars, with gradual
curvature and a high-energy cutoff.   A weak narrow iron feature is also evident in the spectrum (Figure~\ref{fig:fdcut}).

A number of largely empirical models have been developed to describe the broad-band X-ray spectra
of accreting pulsars \citep[see][for a summary of pulsar continuum models]{Muller+13}. In the following, we will apply some of the most common models to our spectrum using the XSPEC software \citep{Arnaud96}. The column density is not 
 constrained by the \nustar\ data alone, and since the low-energy residuals indicate no excess absorption, and past observations did not indicate column densities substantially different from the Galactic value, we fixed its value to the Galactic value $1.010\times10^{22}$cm$^{-2}$ \citep{Kalberla+05}  in all spectral fits \citep[through the {\tt tbabs} model,][]{Wilms+00}. 
 
We first attempted to fit the 3 -- 79\,keV spectrum with a Gaussian iron line plus an absorbed, cutoff powerlaw, where
the continuum model has the form
\begin{equation}
{\tt CutoffPL(E)} \propto E^{-\Gamma} \exp{(-E/E_{\mathrm{fold}})}.
\end{equation}
This model fails to adequately describe the  spectrum, with a best-fit $\chi^2$ of 6864 for 2068 degrees of freedom. We tried to add a Gaussian line around 10\,keV to improve the fit. This ``10\,keV-feature'' is seen in many accreting neutron stars, however, its physical origin is unclear \citep{coburn02a,Muller+13}. The quality of the fit improved clearly to 2482 for 2066 degrees of freedom, when fixing the energy of the feature to exactly 10\,keV (see residuals in Fig.~\ref{fig:fdcut}). However, the best width was around 14\,keV, clearly broader than typically expected for this feature. Leaving the centroid energy free did not result in an improvement of the fit. We therefore rule out a model of a simple cutoff-powerlaw plus a 10\,keV feature.

In order to obtain an adequate spectral fit, we need models with a more flexible continuum shape.   
  We found adequate
 fits using both the Negative and Positive powerlaw Exponential (${\tt NPEX}$, \citealt{Mihara95,1999ApJ...525..978M}), and the
 ``Fermi Dirac Cutoff'' (${\tt FDCUT}$, \citealt{1986LNP...255..198T}) models.   Using the ${\tt FDCUT}$ model, the continuum is
 characterized by
 \begin{equation}
  {\tt FDCUT (E)} \propto A \times E^{-\Gamma} \frac{1}{1 + \exp{((E - E_{\mathrm{cutoff}})/E_{\mathrm{fold}}})}.
  \label{eq:fdcut}
 \end{equation}
 Additionally we applied a Galactic absorption column $N_\text{H}$ and a Gaussian iron line to describe the Fe\,K$\alpha$ line (with energy $E_\text{Fe}$, width $\sigma_\text{Fe}$, and normalization $A_\text{Fe}$). The model is defined in XSPEC as {\tt const*tbabs*(power*fdcut+gauss)}. We obtain a best-fit $\chi^2 = 2266$ for 2067 degrees of freedom.
Figure~\ref{fig:fdcut}
 shows the fit residuals, and Table~\ref{tab:fdcut} shows the best fit parameters with errors.   The residuals clearly
 indicate that above 30\,keV the sharpness of the spectral curvature is not perfectly captured.  
  
 Using the ${\tt NPEX}$ model, with continuum characterized by 
 \begin{equation}
{\tt NPEX (E)} \propto (A_n E^{-\alpha} + A_p E^{+\beta}) \exp{(-E/{kT})},
\label{eq:npex}
 \end{equation}
 we obtain a somewhat better characterization. The negative and positive power-law component have independent normalizations $A_n$ and $A_p$, respecitvely, but are mulitplied by the same exponential turn-over, characterized by the temperature $kT$.  Figure~\ref{fig:fdcut} shows the best fit residuals,
 and Table~\ref{tab:npex}  provides the best-fit parameters. 
The best-fit $\chi^2 = 2234$ for 2066 degrees of freedom. 
  While neither of these models
provides a formally acceptable fit,  the characterization is adequate to understand the basic
continuum and for comparison of the shape with other accreting pulsar
systems.

From the best-fit phase-averaged spectrum using the {\tt NPEX}  model, we find the flux to be $(6.659 \pm 0.001) \times 10^{-9}$\,erg\,s$^{-1}$\,cm$^{-2}$
(3.5 -- 80\,keV).
\citet{Wilson+97} use measurements of the pulsar spin-up during
outburst to place a lower limit on the distance to \gs\ of $d \geq 4.5$\,kpc. \citet{Israel+00} place a higher limit on the distance of 6\,kpc based on considerations on the spectral type of the companion star and the extinction of the source.
Adopting a distance of 5\,kpc implies a 3.5 -- 80\,keV
unabsorbed luminosity of $(2.002 \pm 0.002) \times 10^{37}$\,erg\,s$^{-1}$.

\begin{deluxetable}{rcl}
\tablecolumns{4}
\tablewidth{0pc}
\tabletypesize{\scriptsize}
\tablecaption{Fit parameters for the {\tt FDCUT} model. Parameter names are defined in Eq.~\ref{eq:fdcut}.\label{tab:fdcut}}
\startdata
\hline
$N_{\rm H}$  &=&     $1.010\times 10^{22}$cm$^{-2}$ (fixed)                     \\
$\Gamma$ &=&     $0.64\pm 0.01$  \\
$A$ &=& $\left(9.04\pm0.06\right) \times 10^{-2}$\,ph\,keV$^{-1}$\,cm$^{-2}$\,s$^{-1}$        \\  
$E_\text{cutoff}$ &=& $18.19\pm 0.37$\,keV     \\     
$E_\text{fold} $ &=& $10.10\pm 0.07$\,keV   \\  
$E_\text{Fe} $ &=& $6.51\pm 0.03$\,keV   \\  
$\sigma_\text{Fe} $&=&  $0.27\pm 0.04$\,keV  \\ 
$A_\text{Fe} $&=& $\left(1.06\pm0.10\right)\times10^{-3}$\,ph\,cm$^{-2}$\,s$^{-1}$
\enddata
\tablecomments{All uncertainties refer to single-parameter 90$\%$ confidence limits.  The $\chi^2$ is $2319$ for 2067 degrees of freedom.}
\end{deluxetable}

\begin{deluxetable}{rcl}
\tablecolumns{4}
\tablewidth{0pc}
\tabletypesize{\scriptsize}
\tablecaption{Fit parameters for the {\tt NPEX} model. Parameter names are defined in Eq.~\ref{eq:npex}.\label{tab:npex}}
\startdata
\hline
 $N_{\rm H}$                       &=&      $1.010\times 10^{22}$\,cm$^{-2}$ (fixed)                     \\
$\alpha$              &         =             &    $0.38 \pm 0.05 $     \\
 $\beta$                &     =                 &    $1.50 \pm 0.09 $       \\  
   $kT$                     &    =           &     $7.30 \pm 0.14$\,keV   \\     
   $A_n$                 &= &   $\left(7.75^{+0.35}_{-0.31}\right)\times10^{-2}$\,ph\,keV$^{-1}$\,cm$^{-2}$\,s$^{-1}$\\ 
$A_p$                 &=&     $\left(7.76^{+1.87}_{-1.48}\right)\times10^{-4}$\,ph\,keV$^{-1}$\,cm$^{-2}$\,s$^{-1}$\\ 
$E_\text{Fe} $ &=& $6.51\pm 0.03$\,keV   \\  
$\sigma_\text{Fe} $&=&  $0.27\pm 0.04$\,keV  \\ 
$A_\text{Fe} $&=& $\left(1.07\pm0.11\right)\times10^{-3}$\,ph\,cm$^{-2}$\,s$^{-1}$
\enddata
\tablecomments{All uncertainties refer to single-parameter 90$\%$ confidence limits.  The $\chi^2$ is $2234$ for 2066 degrees of freedom.}
\end{deluxetable}

\section{Phase-resolved Spectral Analysis}\label{sec:phaseres}

\begin{figure}[htbp]
   \centering
  \includegraphics[width=\columnwidth]{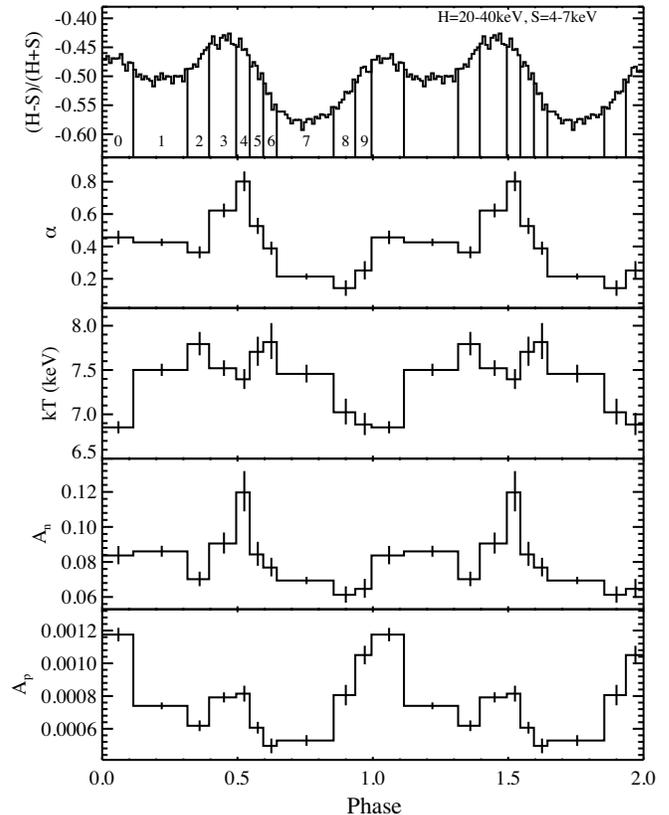} 
   \caption{Hardness ratio between the 20 -- 40 keV and 4 -- 7\,keV bands as a function of pulse
   phase (top panel).  Intervals have been chosen based on evolution of the hardness ration to fit 
   {\tt NPEX} model parameters.  The lower three panels show the evolution of fit parameters as a function
   of pulse phase, i.e., the photon index of the negative powerlaw ($\alpha$), the cutoff energy $kT$, the normalization of the negative powerlaw component $A_n$, and the normalization of the positive powerlaw component $A_p$. The positive powerlaw term is fixed to $\beta$=1.50. The pulse is repeated once for clarity. The error bars are the same as in Table~\ref{tab:npex}.}
   \label{fig:phasebins}
\end{figure}

To further characterize the phase dependence of the spectrum we decomposed the pulse profile into 10 phase bins chosen to 
finely sample  ranges where the hardness ratio is changing rapidly.   We then fit the phase intervals using the
{\tt NPEX} model described above in order to determine the dependence of the model parameters on phase. 
The positive powerlaw term is fixed to $\beta$=1.50,
which is obtained from the phase-averaged spectrum due to the better statistical constraints.
We found that it describes all 10 phase-bin between 3.5--65\,keV very well, with $\chi^2_\text{red}$ values between 0.97 -- 1.06. 
Figure~\ref{fig:phasebins} (top panel) shows the hardness ratio (HR; defined as the ratio of the difference and the sum of the 20--40 and 4--7\,keV fluxes)
as a function of phase. The HR is at a minimum at phase 0.7, roughly coincident with the primary maximum in 
the soft 3 -- 6\,keV profile, and has a maximum 0.85 later in phase, similar to that seen in the raw counts as a function
of phase.  The lower three panels in Figure~\ref{fig:phasebins} show the strong dependence of spectral fit parameters as
a function of pulse phase.    
The two peaks in the hardness ratio, between phases 0.4--0.65 and 0.8--1.1, are originating from different changes in the spectra. The first peak, coincident with the minimum in the soft pulse profiles, has a higher value of $\alpha$. That means, the soft component, described by $\alpha$ and $A_n$, falls off more quickly, resulting in an overall harder spectrum. The second peak in the hardness-ratio, on the other hand, is during the broad peak of the 20--40\,keV pulse profile. Here the change in hardness is mainly due to a drastically increased normalization of the positive power-law component $A_p$, which dominates the spectrum above 
$\sim$20\,keV.

Cyclotron resonance scattering features (CRSFs) are observed in some accreting X-ray pulsars between 5 -- 100\,keV \citep[see][for a review]{Coburn+06}, and yield the only direct measure of the neutron star magnetic field through the Landau relation $E_\text{cyc}=\hbar eB/m_ec= 11.6\,{\rm keV}\,B_{12}$, where $B_{12}$ is the magnetic field in units of $10^{12}$\,G.
In order to search for CRSFs that might provide such measure, we examined the fit residuals in the phase-resolved as well as phase-averaged spectra.   
It has been shown that the exact parameters of a CRSF depend on the chosen continuum \citep[see, e.g.][]{Muller+13}. The NPEX continuum in particular has been found to alter the parameters to unphysical values. However, the NPEX model applied here is smooth in the relevant range for the line ($>$20\,keV), so that a CRSF would show up as a deviation from the continuum. After careful examination of the residuals, we see no systematic features that would indicate the presence of a CRSF in this source.

This negative result of our search could be because  the magnetic field of the source is such that possible cyclotron lines are outside \nustar's bandwidth. This would, however, imply a magnetic field $B \geq 5\times10^{12}$\,G, higher than any confirmed field in a HMXB \citep{caballero12a}. On the other hand, it is possible that the fundamental cyclotron line gets filled up with spawned photons from higher harmonics, making it undetectable in the X-ray spectrum \citep{schoenherr07a}. This scenario has also been discussed for 4U~1909+07, another pulsating HMXB without a detectable CRSF \citep{fuerst12a}. We therefore cannot draw direct conclusions on the magnetic field strength in \gs.

\subsection{Evolution of the Iron Line}

\begin{figure}[htbp]
   \centering
  \includegraphics[width=\columnwidth]{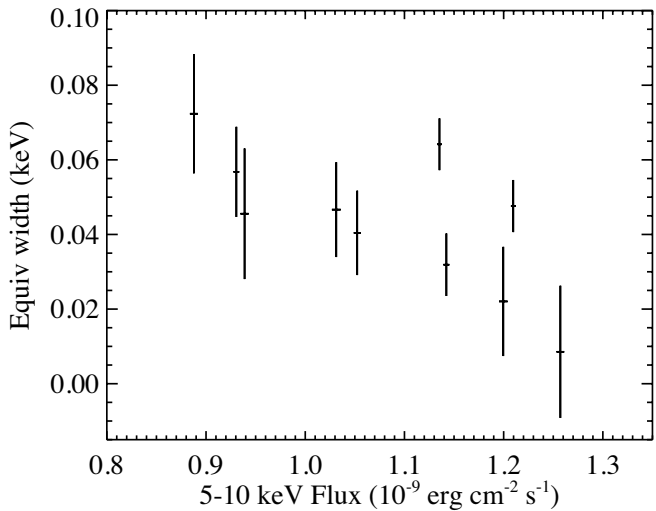} 
   \caption{Scatter plot of iron line equivalent width vs. 5 -- 10\,keV X-ray flux.}
   \label{fig:fevsflux}
\end{figure}

Fe\,K$\alpha$ lines are ubiquitous in spectra of Be/X-ray binaries, and their equivalent widths do not change significantly with luminosity, while their flux is generally proportional to the total flux of the source \citep[see, e.g.,][]{ReigNespoli13}. In these systems they are usually consistent with being produced by fluorescence in cold, quasi-neutral material, probably belonging to the circumstellar material ({\em ibidem}).
We fitted a Gaussian line in every phase of the pulse, in order to study its evolution. Figure~\ref{fig:fevsflux} shows  the equivalent width of the Fe\,K$\alpha$ line, obtained by fitting a Gaussian line with an underlying {\tt NPEX}  continuum model, as a function of soft flux. We observe no significant variation in the line energy, while it appears that the line equivalent width is mildly anti-correlated with the 5 -- 10\,keV flux.
Such a behavior is expected if the line emission is constant with
pulse phase,  and is being diluted by an increased continuum level. This argues in favor of the line produced far from the accretion column, consistent with the general behavior mentioned above. 

\section{Discussion}
\label{sec:discussion}

Time lags between different spectral bands have been observed in accreting systems 
ranging from low-mass X-ray binaries to active galactic nuclei.  A number of processes can produce these lags, such as: the reprocessing of a source signal through inverse Compton scattering \citep{SunyaevTitarchuk80,Payne80,Liang+84};
reflection from an accretion disk or companion star \citep{Basko+74,GeorgeFabian91};  or
disk wave propagation \citep{Lyubarskii97,Kotov+01,ArevaloUttley06}. See also \citet{Poutanen01} for a review. 
Recently, sophisticated methods have been developed to relate spectral and timing variability 
using time lags to identify underlying processes or constrain the geometry of the emission \citep{Casella+04,Reig+06,Uttley+11,Cassatella+12,Barret13}. 
Many other accreting X-ray pulsars show time lags in their pulsed emission \citep[e.g.][]{Cui+98,Ibragimov+11,Falanga+12}, 
but they appear quite different from what we observe in \gs.  They are usually seen in the soft band (E$<5$\,keV) and are of much lower magnitude ($<0.05$ cycles in phase). 
In this section we discuss possible origins of the large phase shifts seen up to 80\,keV by \nustar.

\paragraph{Reflection from the accretion disk}
Reflection of an X-ray signal off an accretion disk produces a very specific hardening of the spectrum \citep{Basko+74,GeorgeFabian91}. 
Lags are produced by the different path lengths and therefore light travel time of the original and the reflected signals. 
The far edges of an accretion disk of radius $R$ are expected to be geometrically thick, with the scale height $H$ reaching up to $H/R\sim 0.5$ \citep{WhiteHolt82}, 
and therefore the reflecting region can be quite far from the central source.   In a binary, the X-ray emission can also be reflected off a dense
stellar wind or the companion star.  The energy dependency of the lags could possibly arise from the combined effect of the 
drop in the source signal and the rise of the reflected signal at different phases of the pulse.

However, we find this mechanism to be an unlikely explanation for the lags seen in \gs.  
First of all, a reflection model requires fine tuning to produce the smooth phase shift seen in Figure~\ref{fig:pulsemap}. 
In order to obtain a sharp reflected pulse, there would have to be only a narrow region of the accretion disk intersected by the pulsar beam, 
otherwise the reflected pulse would be spread by an amount proportional to the size of illuminated region. We note, though, that a geometry involving a warped disk might decrease the spread of the pulse \citep[see, e.g.,][]{HickoxVrtilek05}. 
The large orbit of this system ($a_x \sin i \sim 130$\,lt-s, see Table~\ref{tab:ephemeris}) also rules out reflection from the surface of the companion star \citep[see, e.g.,][]{Chester79}, due to inconsistent timescales. 

As a final remark, the very high reflected fraction that would be necessary to explain the hardening of the spectrum observed between intervals 7 -- 9 in Figures~\ref{fig:countrate} and \ref{fig:phasebins} (we use the XSPEC model {\tt reflect} to estimate that a covering
fraction of $\sim$0.7 would be required) is inconsistent with the very low equivalent width of the iron line \citep{GeorgeFabian91}. The evolution of these phase-resolved spectra is instead well described by models commonly used to fit high-mass X-ray binary spectra.
In such systems, the hard emission arises from soft photons being
Compton up scattered in the accretion stream \citep{BeckerWolff07}.

\paragraph{Comptonization by a Corona}

Inverse Compton scattering of soft X-rays off a hot electron corona that surrounds the compact object and the disk is an 
alternative way to produce hard lags, and it has been proposed as an explanation in some 
black hole binaries \citep[e.g.][]{Liang+84,Miyamoto+88,Dove+98,LeeMiller98,Reig+00}. 
The scattering that hardens the spectrum also introduces a delay in the reprocessed emission. 
If one assumes that the temperature of the corona is constant, that each scattering produces the 
same energy transfer, and that the flight time of the photons $\delta$ is the same between every scatter and 
equal to the mean length between scatters in light seconds $r/c$, it is easy to show that the total delay $\Delta$ introduced by the process is \citep{Nowak+99}
\begin{equation}\label{eq:compt}
\Delta \simeq n \delta \simeq \dfrac{r}{c} \dfrac{mc^2}{4kT_e} \ln{\dfrac{E_n}{E_0}} 
\end{equation}
\\
where $n$ is the number of scatters to go from energy $E_0$ to $E_n$, $mc^2$ the electron rest mass, and $kT_e$ the temperature of the electrons.
By fitting the data in Figure~\ref{fig:lagfit} with the formula in Eq.~\ref{eq:compt}, and assuming an electron temperature of $50$\,keV, as it is often observed in accreting sources, we obtain a mean flight length of $r/c\sim 0.5$\,s, 
which is not unlikely in itself, given that in HMXBs a large shocked region is expected around the neutron star, where the electron temperature can reach very high values \citep{Nagae+04, Mauche+07}.

This process has the advantage of smoothly relating the magnitude of the lag to the energy of the reprocessed photons. The main problem in this scenario is that Comptonized photons lose directional information after the first scatter. 
This means that, to zeroth order, we would expect reprocessed photons coming from the different phases of the pulsar rotation to reach the observer at the same moment. Some work on a similar model has been done in the past for Her X-1 \citep{Kuster+05} and the result is that with this process no sharp information can survive. 

This scenario is therefore unsuitable for explaining the very sharp shape of the lagged pulses, 
unless the Comptonizing region is not distributed uniformly around the source, but is instead concentrated in a small region along the line of sight and is thus only illuminated in a well-determined pulse phase.

\paragraph{Pulsar Beam Shape-Induced Delays}

In the accretion column of accreting X-ray pulsars the combined effect of ram pressure from the in-falling matter, 
radiation, magnetic field and gravitational effects can give rise to a complex beam structure with evolving spectral states driven mainly by accretion rate \citep[e.g.][]{BlumKraus00,Leahy04a,Leahy04b,Becker+12}. 
This complex structure is likely to show different spectra in different regions of the beam, as often shown through phase-resolved spectroscopy.

Accreting X-ray pulsars often show phase/time delays in their pulsed emission. 
There is a wide range of examples which appear phenomenologically different, suggesting that a single model does not apply to all sources. 
Soft lags, as opposed to the hard lags reported here, are quite common in accreting millisecond pulsars, like in SAXJ1808.4$-$3658 \citep{Cui+98}, IGR J17511$-$3057 \citep{Ibragimov+11}, or IGR J17498$-$2921 \citep{Falanga+12}, 
but their magnitude is below the $\sim$ms level. They can be modeled as a 
Relativistic Doppler effect (the beam appearing harder when it is moving towards the observer) or 
Compton down scatter in the accretion column. Similar behavior has been observed in burst oscillations  \citep{Ford99,Artigue+13}. 
More peculiar behavior has been observed in the Be/X-ray binary 4U 0115+63 \citep{Ferrigno+11}, with unusual ``waves" in the 
pulse phase lags that the authors convincingly attribute to an effect of cyclotron resonant scattering. 

If the source has a luminosity around $\sim10^{37}$\,erg\,s$^{-1}$, as is the case for \gs\ in outburst, latest calculations show that a Compton shock forms in the accretion column, but that the luminosity is not sufficient to produce a radiation dominated shock, which would completley decelerate the infalling material high above the neutron star surface \citep{Becker+12}. In this intermediate regime, a complex beam profile is present, with radiation emerging through the walls of the accretion column (fan beam), as well as along the magntic field lines (pencil beam), after being upscattered to high energies by inverse Compton scattering inside the accretion column.
The fan and pencil beam components could in principle have different spectral properties and, given the right geometry, could create the phase-shift effect seen in the \nustar\ data
through a gradual mixing of a softer and harder continuum.

The situation is even more complex as relativistic light bending can change the observed spectrum with pulse phase 
quite drastically, depending on the altitude above the neutron star \citep{Kraus01}. 
Furthermore a filled accretion funnel is only the simplest geometry; strong evidence exists that the accretion column 
might be hollow for certain sources, with different spectra emerging from the inner and outer walls 
(A 0535+26, \citealt{Caballero+11}; Cen X$-$3, \citealt{Kraus+03}). 
These complex geometries could lead to the observed pulse profile shift in \gs. 

Pulsar beams are known to be  complex and diverse,  while Comptonization and reflection are 
ubiquitous phenomena and corona and disk geometries are likely similar in different sources.
We would therefore expect hard lags to appear in more systems were they caused by the latter effects. 

\section{Conclusions}

We have analyzed observations of the accreting X-ray pulsar \gs\ using the \nustar\ high-energy X-ray telescope. 
The spectrum appears to be characteristic of those observed in HMXBs, and is well described by the 
phenomenological models {\tt NPEX} and {\tt FDCUT} with parameter values within the ranges typically used to describe HMXB spectra.  
We detect a weak iron  line (EW $\sim$  40\,eV) which appears mildly anti-correlated with the 5 -- 10\,keV flux.    
We find no evidence for a cyclotron resonance scattering feature in either the phase-averaged or phase-resolved spectra.

The pulse profile evolves strongly with energy.   The 3 -- 10\,keV profile has two peaks; a secondary peak at phase 0.2 (relative to an
arbitrary reference time) and a primary peak at phase 0.8. 

We observe a strong phase shift of these peaks with energy, up to almost 0.3 (about 4\,s in time)  in phase. The shift  is almost linear at low energies,  saturating 
above $\sim$40\,keV.    A significant phase shift of this magnitude
at hard X-ray energies has not been observed before in an accreting X-ray pulsar.   This shift is hard to explain as a result either of
time lags due to reflection or Comptonization.   
The unusual spectral evolution with phase might be instead related to the detailed structure in the pulsar beam.
In this case, our results will help shed light on the nature of the physical effects taking place in the pulsar accretion column.

\acknowledgments
This work was supported under NASA Contract No. NNG08FD60C, and
made use of data from the {\it NuSTAR} mission, a project led by
the California Institute of Technology, managed by the Jet Propulsion
Laboratory, and funded by the National Aeronautics and Space
Administration. We thank the {\it NuSTAR} Operations, Software and
Calibration teams for support with the execution and analysis of
these observations.  This research has made use of the {\it NuSTAR}
Data Analysis Software (NuSTARDAS) jointly developed by the ASI
Science Data Center (ASDC, Italy) and the California Institute of
Technology (USA).  Matteo Bachetti wishes to acknowledge the support 
from the Centre national d'\'etudes spatiales (CNES).
Lorenzo Natalucci acknowledges financial support through contract 
ASI/INAF I/037/12/0.


\end{document}